\documentclass[12pt,preprint2]{aastex}
\usepackage{epstopdf}

\shortauthors{Mitchell}

\begin{document}

\title{Coupling convectively driven atmospheric circulation to surface rotation:  Evidence for active methane weather in the observed spin rate drift of Titan}

\author{Jonathan L. Mitchell\altaffilmark{1}}
\affil{Institute for Advanced Study, Princeton, NJ, 08540}
\email{mitch@ias.edu}

\begin{abstract}
A large drift in the rotation rate of Titan observed by Cassini provided the first evidence of a subsurface ocean isolating the massive core from the icy crust.  Seasonal exchange of angular momentum between the surface and atmosphere accounts for the magnitude of the effect, but observations lag the expected signal by a few years.  We argue this time lag is due to the presence of an active methane weather cycle in the atmosphere.  An analytic model of the seasonal cycle of atmospheric angular momentum is developed and compared to time-dependent simulations of Titan's atmosphere with and without methane thermodynamics.  The disappearance of clouds at the summer pole suggests the drift rate has already switched direction, signaling the change in season from solstice to equinox.  
\end{abstract}


\section{Introduction}
Models of the interior of Titan for plausible ammonia concentrations predict the existence of a water ocean under a thin ice I crust of perhaps 100 km thickness \citep{2003JGRE..108.5130S}.  A subsurface ocean would dynamically isolate the surface from the more massive core, effectively lowering the moment of inertia of the surface.  For the expected magnitude of seasonal angular momentum change in the atmosphere, a thin ice crust experiences measurable changes in length-of-day, or equivalently spin rate, in a matter of a few years \citep{2005GeoRL..3224203T}.  

Cassini RADAR spin rate measurements \citep{2008AJ....135.1669S}, seasonal winds from an atmospheric model prediction \citep{2005GeoRL..3224203T} and Huygens wind observations \citep{Bird:2005ai} were used to infer the moment-of-inertia (MOI) of Titan's surface \citep{2008Sci...319.1649L}.  The surface MOI is well below that allowed for any solid internal configuration, but is consistent with models of Titan's interior that with a thin surface shell overlaying a subsurface ocean \citep{2003JGRE..108.5130S}.  Seasonally changing zonal winds in the atmosphere torque the freewheeling surface, introducing a forced libration about synchronous rotation.  The forced libration of the surface, however, unexpectedly lags seasonal forcing by a few years.  Here we suggest this time lag signals an active methane weather cycle in the atmosphere.   

Titan's atmosphere is known to superrotate, with angular momentum exceeding that of solid body rotation.  Stratospheric zonal winds exceed 100 m/s in the prograde direction \citep{1993Icar..103..215H, 2008Icar..194..263A}.  Between 10 and 60 km, the troposphere weakly superrotates with winds limited to 10-20 m/s \citep{Bird:2005ai}.  The lowest $\sim$10 km at low latitudes show slight retrograde motion, which is consistent with a seasonal, cross-equatorial circulation, or Hadley cell, driven by convection \citep{2007P&SS...55.1990T}.  Winds in the lowest kilometer turn prograde again \citep{2007P&SS...55.1896K}, which is corroborated by the directionality of equatorial dune features \citep{Lorenz:2006fv}.\footnote{The presence of prograde winds at the equatorial surface is mysterious given frictional coupling with the surface.  Explanation of this phenomenon is beyond the scope of this work.}  Atmospheric models suggest angular momentum is seasonally redistributed in the atmosphere by a cross-equatorial circulation that changes direction with seasons \citep{Hourdin.95, Tokano:1999fk, Tokano.et.al.05, Rannou.et.al.06, Mitchell:2006lr, 2007P&SS...55.1990T, Mitchell:2008}. 
 
Though stratospheric zonal winds dwarf those in the troposphere, very little mass is being advected in the stratosphere.  Zonal winds in more than 90\% of the atmospheric mass are below 20 m/s.  The lowest one-third of the atmospheric mass is subject to the largest seasonal change in angular momentum due to frictional torque from the surface acting on a seasonally reversing Hadley cell.  In the following, we develop a theory for the magnitude of seasonal change in atmospheric angular momentum (AAM) and its effect on the spin rate.  We then develop a full time-dependent model of this process and show by comparison with numerical simulations that a seasonal time lag of a few years is a natural consequence of methane thermodynamics in Titan's atmosphere.  We conclude by offering predictions for current and future Cassini observations. 

\section{A model for seasonal atmosphere-surface angular momentum exchange}
Here and throughout we assume axisymmetry of the atmosphere; by doing so we preclude the development of superrotation.  Asymmetric surface heating by sunlight drives a convective updraft at a certain latitude.  Surface friction imparts angular momentum to converging, surface-level air, which is then lofted into the upper troposphere by the convective updraft.  At the top of the convective layer, air diverges along meridians while conserving angular momentum.  The Hadley cell homogenizes the angular momentum in this upper, divergent branch of the flow.  The triple requirements of angular momentum and energy conservation and cyclostrophic balance \citep{2006aofd.book.....V} allow prediction of the width and strength of the Hadley circulation \citep{Held.and.Hou.80}; this model predicts a global Hadley cell on Titan.  Air sinks at the poles and returns along the surface where friction resets the angular momentum to the surface value.  

\subsection{Atmospheric angular momentum}
In our model, the maximum achievable specific angular momentum at any location in the Hadley domain, i.e.~the convective layer of the atmosphere, is that of the equatorial surface
\begin{equation}
l_{max} \ = \ \Omega a^2
\label{eq:Mmax}
\end{equation}
where $\Omega$ is the rotation rate and $a$ is the planetary radius.  The angular momentum of the updraft is nearly zero when the convective updraft is located at the poles.  Provided the updraft reaches the poles at or near solstices, the maximum change in specific angular momentum is equal to the maximum, Equation \ref{eq:Mmax}.  If we further assume the Hadley cell distributes this angular momentum evenly over a layer of thickness $\Delta p_a$, the maximum change in atmospheric angular momentum is,

\begin{equation}
\Delta L_{atm} \ = \ M_a \Omega a^2 
\label{eq:deltaLatm}
\end{equation}
where 
\begin{equation}
M_a \ = \ 4 \pi a^2 \Delta p_a/g
\end{equation}
is the mass of the layer where the Hadley cell homogenizes angular momentum (assuming hydrostasy).  

\subsection{Effect on spin rate}
Conservation of angular momentum dictates that the surface spin rate must change in response to changes in atmospheric angular momentum.  Seasonal changes in atmospheric angular momentum can measurably alter the spin rate of the surface, particularly if there exists a subsurface ocean isolating the icy lithosphere from the core \citep{2005GeoRL..3224203T,2008Sci...319.1649L}.  The moment of inertia for a thin, isolated shell of thickness $\Delta r_i$ is
\begin{equation}
\label{eq:Is}
I_s 
\simeq  \frac{2}{3} M_i a^2 \ \nonumber
\end{equation}
where 
\begin{equation}
M_i \ \simeq \ 4 \pi a^2 \rho_i \Delta r_i
\end{equation}
is the mass of the shell and $\rho_i$ is the density of ice I.  The maximum seasonal change in the spin rate of this shell is
\begin{eqnarray}
\Delta \Omega_{max} & = & -\Delta L_{atm}/I_s \\	 \nonumber
				  & \simeq & \frac{3}{2} \frac{M_a}{M_i} \ \Omega \ .
\label{eq:deltaOm}
\end{eqnarray}

\subsection{Maximum angular momentum exchange and spin rate drift}
The convective layer on Titan extends from the surface (1500 mbar) to roughly 1000 mbar.  If the upper branch of the Hadley cell extends through the top half of this layer, $\Delta p_a$ = 250 mbar; taking g = 1.35 m/s$^2$, a = 2575 km, and $\Omega$ = 4.5 $\times$ 10$^{-6}$, the maximum seasonal change in atmospheric angular momentum in the Hadley domain is $\Delta L_{atm} \sim$ 2.5 $\times$ 10$^{25}$ kg m$^2$/s.  This value is comparable to the change predicted in a full general circulation model (GCM) \citep{2005GeoRL..3224203T}, suggesting the seasonal cycle of the Hadley cell is dominating changes in the atmospheric angular momentum.  If we assume $\rho_i$ = 917 kg/m$^2$, $\Delta r_i$ = 70 km and $\Delta p_a$ = 150 mbar, then $M_a/M_i \ \simeq \ 4 \times 10^{-4}$ and the maximum seasonal spin rate drift is $\Delta \Omega_{max} \sim 2$ deg/year, which is again in rough correspondence with the GCM prediction \citep{2008Sci...319.1649L}.

\subsection{Time-dependent angular momentum exchange}
The arguments presented so far can be generalized into a time-dependent prediction of spin rate change for comparison with observations.  We make the following assumptions: 1) the convective updraft of the Hadley cell follows in-phase with the latitude of maximum diurnal-average solar forcing, $\varphi_1$ and 2) the Hadley cell instantaneously resets atmospheric angular momentum to the value of the surface at $\varphi_1$.  With these assumptions, the time-dependent atmospheric angular momentum is
\begin{equation}
L_{atm}(\varphi_1(t)) \ = \ Const. + \frac{1}{2} M_a \ \Omega a^2 \cos^2(\varphi_1(t))
\end{equation}
where the constant results from the planetary and stratospheric angular momenta that we assume change very little.  All of the time dependence enters through the latitude of the updraft, $\varphi_1$. 

\begin{figure*}[htbp]
\begin{center}
\plotone{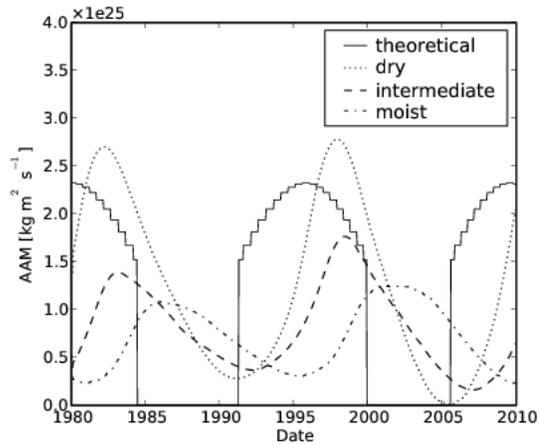}
\caption{Relative atmospheric angular momentum (AAM) over one Titan season for our analytic and numerical models.  See text for descriptions.}
\label{fig1}
\end{center}
\end{figure*}

Figure \ref{fig1} shows the time-dependent model prediction of atmospheric angular momentum for previously chosen parameters with a constant subtracted (solid line).  The latitude of maximum diurnally averaged solar forcing discontinuously switches from mid summer latitudes to the summer pole near solstices, during which time the model atmospheric angular momentum is a minimum.  Figure \ref{fig2} shows the model prediction of spin rate drift (with the average subtracted) assuming the surface is a 70 km thick ice I shell (solid line).  The bold line shows the measured spin rate drift \citep{2008AJ....135.1669S}.  The model clearly predicts the correct magnitude, but fails to produce the observed time lag.  

\section{The role of methane thermodynamics in producing a time lag}
The updraft of the Hadley cell is located where convection is favorable, which on Earth is nearly steadily located at warm equatorial oceans.  On Titan, the surface warms quickly relative to a season, and one therefore might expect the convective updraft to follow the latitude of maximum solar forcing.  However, the atmosphere could not quickly respond to such warming since the radiative cooling time exceeds the length of a Titan year \citep{FLASAR:1981lr}.  For reasonable estimates of surface heat capacity \citep{Tokano.et.al.05}, the radiative cooling time of the surface is perhaps a month; it is clear that the atmosphere provides the dominant source of heat capacity in the climate system.  

Dry (i.e.~neglecting methane condensation and evaporation) GCM simulations suggest the large heat capacity of the atmosphere does not significantly attenuate the seasonal cycle of the Hadley cell; the convective updraft of the Hadley cell travels from one pole to the other each Titan year, roughly in phase with the seasonal cycle of solar radiation \citep{Tokano.et.al.05}.  However, increasing the amount of methane evaporation at the surface introduces a time lag of the response of the Hadley circulation to seasonal forcing \citep{Mitchell:2006lr}.  Moist, precipitating convection penetrates deeper into the cold, dense atmosphere than dry convection; one can think of the increase in convective depth arising from a reduction in the vertical temperature gradient resulting from moist convection versus dry convection (the lapse rate).  By doing so, moist convection communicates surface warming into the large heat capacity reservoir in the upper troposphere that is not accessed by dry convection alone.  The Hadley cell efficiently communicates convective warming throughout the latitudinal domain, which increases the thermal inertia of the system.  The net result is an increase in the mass of the atmosphere through which the warming of the subsolar surface is communicated, which introduces a time lag in the response of the Hadley cell to seasonal forcing.

Figure \ref{fig3} shows the predicted vertical pressure velocity with the scale truncated to highlight regions of large-scale updraft (colored contours) as a function of latitude and date over the last half Titan year of three model simulations\footnote{This model is similar to our earlier one \citep{Mitchell:2006lr} except a more realistic radiation scheme was implemented \citep{Mitchell:2008}.}; the insolation pattern is overplotted for reference (black contours).  The position of the most intense updraft features in Figure \ref{fig3} relative to the solar forcing gives a measure of the time lag of the Hadley cell.  The ``dry'' limit resembles other dry simulations \citep{Tokano.et.al.05}, and very little time lag is predicted \citep{2008Sci...319.1649L}.  In the ``moist'' case, in which deep, precipitating convection is most favorable, the main convective updraft is in quadrature with the seasonal forcing.  The ``intermediate'' case, in which methane evaporation was restricted to produce $\sim$50\% relative humidities near the surface, has sporadic precipitating convection, and as a result has a smaller time lag. 

Atmospheric angular momentum (with the minimum subtracted) and the induced spin rate drift (for an assumed 70 km ice I shell with the average drift subtracted) over one Titan year for our three simulated cases are shown in Figures \ref{fig1} and \ref{fig2}, respectively.  The time lag of the Hadley cell updraft evident in Figure \ref{fig3} translates directly into a time lag in these quantities.  For comparison, the measured spin rate drift and the acceleration are shown as a solid, bold line \citep{2008Sci...319.1649L, 2008AJ....135.1669S}.  The ``intermediate'' case fits the data very well, indicating methane thermodynamics plays a role in establishing the observed time lag in the measured spin rate drift.  While our dry model reproduces the change in atmospheric angular momentum in the previous model estimate \citep{2005GeoRL..3224203T}, the change is somewhat smaller in the intermediate and moist cases.  Our model predicts the spin rate drift should currently be changing directions; Cassini RADAR observations through 2010 should unambiguously determine whether our prediction holds.  

\section{Thermodynamic efficiency of Hadley cell oscillations:  Comparison with Earth}
Work must be done to change the spin rate of the ice shell.  The torque exerted by Hadley cell oscillations is driven by seasonal changes in insolation.  The thermodynamic efficiency of this process can be estimated given the change in rotational kinetic energy and the integrated sunlight over a quarter Titan season.  

The kinetic energy of the surface shell is
\begin{equation}
T_s \ = \ \frac{1}{2} I_s \Omega^2 \ .
\end{equation}
The change in kinetic energy resulting from a change in surface angular momentum, $\Delta L = I_s \Delta \Omega$, is approximately
\begin{eqnarray}
\label{eq:Tsmax}
\Delta T_{s} & \simeq & \Delta L \Omega \ .
\end{eqnarray}
For the numbers listed previously, $\Delta T_s \sim 10^{20}$ Joules.  The time to go from maximum to minimum spin is roughly a single season, or $\sim$7 years; in this time, Titan receives 7 $\times 10^{22}$ J of radiant energy from the Sun.  The thermodynamic efficiency of seasonal angular momentum exchange is $\epsilon \sim 0.1$\%.  \citet{2003JGRE..108.5130S} suggest tidal dissipation in Titan could be 10$^{10}$ W which over a Titan season amounts to $\sim 10^{18}$ Joules, two orders of magnitude smaller than the surface kinetic energy change.  

Earth experiences seasonal changes in length of day of a millisecond due to atmospheric torques.  Earth's moment of inertia about the spin axis is 8 $\times 10^{37}$ kg m$^2$, and the kinetic energy associated with the spin rate change is $\sim 5 \times 10^{21}$ Joules (see \citet{1993SGeo...14....1R} for a review of the topic).  Earth receives $\sim10^{24}$ Joules of radiant energy from the Sun in a season, implying an efficiency of $\epsilon \sim 0.5$\%, the same magnitude as that of Titan.  Titan receives 1/15th of the solar energy the Earth receives over the time of their respective seasons (7 years for Titan, 3 months for Earth).  Taking the change in AAM as a measured quantity, the difference in solar energy nearly cancels the factor of 16 due to the difference in rotation rate in Equation \ref{eq:Tsmax} when estimating the efficiency.  The change in Titan's relative AAM, $L_{max}$, is $\sim$1/5th of Earth's, which accounts for Titan's lower efficiency.  However, since Titan's stratosphere either absorbs or reflects more than half of the incoming solar radiation \citep{MCKAY:1991fu}, Titan's tropospheric Hadley cell is driven by a reduced level of solar radiation relative to the top-of-atmosphere.  If this is taken into account, the efficiency increases to near the level of Earth.  

\citet{1993SGeo...14....1R} attributes Earth's seasonal changes in length-of-day to variability in the subtropical jets, which mark the poleward flanks of Earth's angular-momentum-transporting Hadley cell.  Figure \ref{fig4} compares changes in relative AAM on Earth from NCEP/NCAR reanalysis data for 2007 \citep{1996BAMS...77..437K} to the our ``intermediate'' Titan simulation; the top two panels display the vertical mass-weighted average of $u a \cos(\varphi)$, which is the specific angular momentum of the atmosphere relative to rotation (measured positive in the direction of rotation), and the bottom two panels show the corresponding relative AAM.  Earth's subtropical jets contribute bands of positive AAM between 30-40 N/S latitude in Figure \ref{fig4}(a).  The Northern hemisphere jet has stronger seasonality, with a minimum during Northern hemisphere summer corresponding to a global minimum of relative AAM in Figure \ref{fig4}(c).  Titan's subtropical jets are spread further poleward because of the increased width of the Hadley cell (Figure \ref{fig4}(b)), but the pattern with season shows similarity to the Earth's.  In particular, minima in relative AAM occur when either the Northern or Southern subtropical jet is at minimum.

\section{Other potential sources of time lag}
There are a few physical processes which could additionally contribute a time lag, and we presently estimate their effects.  First, a sluggish atmospheric overturning circulation might contribute a time lag; we assumed in our analytic model that the circulation is vigorous and able to redistribute the angular momentum of the convective updraft instantaneously.  The overturning timescale varies by an order of magnitude between the ``moist'' case and the ``dry'' limit.  Since our analytic model assumes the atmosphere instantaneously acquires the angular momentum of the surface at the updraft, the effect of a sluggish circulation is likely positively contributing to the time lag in AAM.  Since the atmosphere is most sluggish in the ``moist'' case, the effect goes in the same direction as the thermal inertia.  The weaker Hadley circulations in the ``intermediate'' and ``moist'' cases reduce the amplitude of AAM changes relative to the ``dry'' limit, as seen in Figure 1 in the main text; as a result, we infer a somewhat shallower ice shell depth of $\sim$70 km.  

Second, frictional (Ekman) drag at the interfaces of the subsurface ocean could dynamically couple the surface to the deep core.  Though the drag is only communicated to a thin layer of depth $\delta \sim (2 \nu/f)$, where $\nu$ is the viscosity of water and $f \sim 2 \Omega$ is the coriolis parameter, mass transport in this layer converges (diverges) to form downwelling (upwelling) through the depth.  The characteristic timescale of this overturning is $\tau_{Ek} \sim H (\frac{1}{2} f \nu)^{-1/2}$, where $H$ is the ocean depth \citep{2006aofd.book.....V}.  \citet{2003JGRE..108.5130S} estimate the ocean depth to be 224 km for an assumed 15 wt.\% NH$_3$.  If we assume $H \sim$ 200 km and $f \sim 2 \Omega$, then $\tau_{Ek} \sim$ 3000 years.   Ekman drag in a laminar boundary layer cannot contribute a time lag of a few years.  However, the Richardson number, $Ri$, in an assumed laminar boundary layer may be order unity\footnote{We estimate the interior heat flux from the model of \citet{2003JGRE..108.5130S} and calculate the temperature difference over the laminar Ekman layer due to thermal conduction.  We then use this temperature difference along with an estimate of the velocity in the boundary layer, $u \sim \Delta \Omega a$, to estimate $Ri \sim 0.7$}`.  In this case the molecular viscosity, $\nu$, should be replaced with an effective eddy viscosity, $\kappa$.  We estimate $\kappa \sim L_d f/\sqrt{Ri}$, where $L_d \sim u/f$ is the radius of deformation \citep{2006aofd.book.....V} and $u \sim \Delta \Omega a$ is the velocity difference in the boundary layer induced by libration of the crust.  For previously listed values, $\kappa \sim 2-3 \times 10^{-2}$ m$^2$/s and the Ekman spin-down due to the turbulent boundary layer is $\tau_{Ek} \sim$ 20 years.  This rough estimate suggests that if the boundary layer is turbulent, dynamic coupling of the surface and core by Ekman drag could introduce a time lag in the drift rate.  

Third, we have implicitly assumed torques on a permanent quadrupole moment in the crust cannot overcome frictional torques provided by the atmosphere, which may not be the case.  The full equation of motion (EOM) of the surface crust including these terms reads
\begin{equation}
C_s \ddot{\Theta} \ = \ T_{atm} + T_{Sat} + T_{coup} \ ,
\label{eq:EOM}
\end{equation}
where $C_s = I_s$ is the polar moment of inertia of the crust and $\ddot{\Theta} =$ 0.1825 deg/yr$^2$ is the measured rotational acceleration of the surface.  The $T$'s represent torques of various mechanisms averaged over an orbital period.  $T_{atm}$ is the seasonal torque of the atmosphere; our analysis has assumed this term dominates the others, $C_s \ddot{\Theta} \simeq T_{atm}$.  $T_{Sat}$ is the torque of Saturn's gravitational field acting on permanent deformations contributing a quadurpole moment to the crustal layer.  $T_{coup}$ represents gravitational coupling between a permanent surface quadrupole and a permanent quadrupole in the core.  Gravity measurements from the Cassini spacecraft can constrain the quadrupole moments, but none have yet been reported in the literature.  By assuming they do not contribute significantly to Equation \ref{eq:EOM}, we have implicitly placed limits on the surface and core quadrupoles contributing to $T_{Sat}$ and $T_{coup}$, which we presently quantify.  We will express limits in terms of the ratio $(B-A)/C$, where $C$ is the polar moment of inertia and $B$ and $A$ are the moments orthogonal to the spin axis.  The subscripts ``$s$'' and ``$c$'' will respectively refer to moments of the surface and core.  

Saturn's gravity would apply a torque to a permanent quadrupole in the surface.  Neglecting this effect in the EOM implies $\ddot{\Theta} >> T_{Sat}/C_s$.  The torque is proportional to $(B-A)_s$, the quadrupole of the surface, and $n \simeq \Omega$, the orbital mean motion \citep{murray.dermott}.  Together with the Cassini measurement of rotational acceleration, the constraint implies $((B-A)/C)_s << 10^{-7}$.  Assuming the surface is a dynamically isolated ice shell of 70 km thickness, $(B-A)_s << 2 \times 10^{27}$ kg m$^2$.  The maximum $(B-A)_s$ could, for instance, be due to an isostatically uncompensated disk of ice of depth 1 km and diameter smaller than 40 km sitting on the equatorial surface; a 100 m thick disk would need to be smaller than 125 km.

A surface quadrupole could also gravitationally couple to a quadrupole in the core, represented by $(B-A)_c$.  As expressed in the EOM, we have assumed $\ddot{\Theta} >> T_{coup}/C_s$.  We derive a constraint on $((B-A)/C)_c$ by making the following assumptions:  1) the quadrupole of the surface layer relative to its polar moment takes the maximum value estimated in the previous paragraph, $((B-A)/C)_s \sim 10^{-7}$; 2) the core remains fixed in the rotating frame; and 3) the core radius is 2250 km and has the composition shown in Figure 6 of \citet{2003JGRE..108.5130S}.  Under these assumptions, $((B-A)/C)_c << 5\times 10^{-5}$.  This minimum value is intermediate between the moon (which is roughly the size of Titan's rock and iron core), $((B-A)/C)_{m} \sim 2 \times 10^{-4}$, and the Earth, $((B-A)/C)_{\earth} \sim 2 \times 10^{-5}$.  

\section{Conclusions}
\begin{figure*}[htbp]
\begin{center}
\plotone{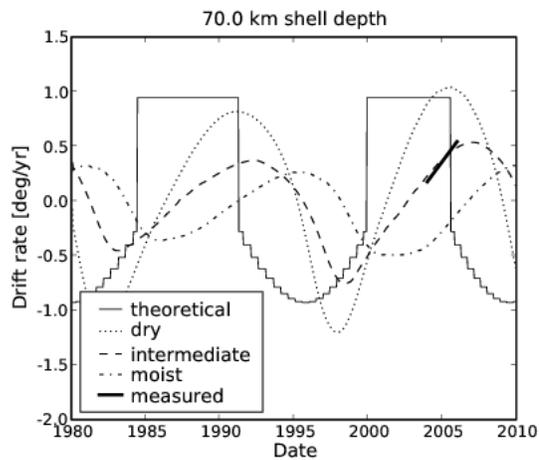}
\caption{Induced spin rate drift over one Titan season assuming a uniform, dynamically isolated ice I shell of 70 km thickness.  Lines are defined as in Figure \ref{fig1}.}
\label{fig2}
\end{center}
\end{figure*}

\begin{figure*}[htbp]
\begin{center}
\plotone{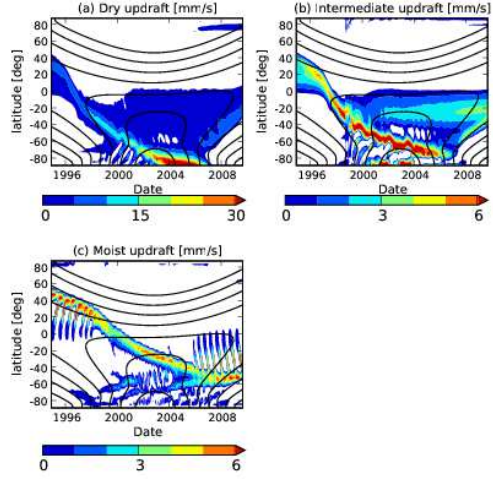}
\caption{Updraft velocity at the lowest model layer in mm/s (colored contours) and solar forcing (black contours) during the Cassini epoch for our numerical simulations.  See text for definitions of each case.}
\label{fig3}
\end{center}
\end{figure*}

\begin{figure*}[htbp]
\begin{center}
\plotone{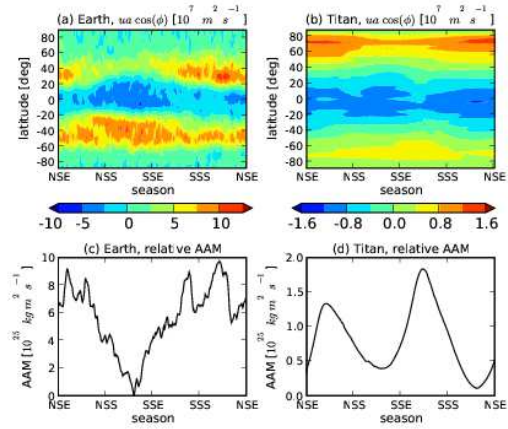}
\caption{Vertical (mass-weighted) average of specific relative atmospheric angular momentum of Earth (a) and Titan (b) as a function of latitude over one full seasonal cycle.  N/SSE -- Northern/Southern Spring Equinox; N/SSS -- Northern/Southern Summer Solstice.  Global relative atmospheric angular momentum of Earth (c) and Titan (d).  Earth diagnostics performed on NCEP/NCAR reanalysis data for 2007 \citep{1996BAMS...77..437K}.  Titan diagnostics performed on the ``intermediate'' simulation.}
\label{fig4}
\end{center}
\end{figure*}

We have developed an analytical model for the time-dependent atmospheric angular momentum in an atmosphere with an oscillating Hadley cell which agrees with a simulation of Titan's atmosphere \citep{2005GeoRL..3224203T}; while both models predict the observed magnitude of spin rate drift, neither can produce the observed time lag of the seasonal drift.  Numerical simulations with methane thermodynamics included show an increase in the thermal inertia of the system, thus introducing a time lag with the seasonal forcing.  The time lag is sensitive to the supply of methane evaporation at the surface.  If methane is allowed to efficiently evaporate from the model surface, deep precipitating convection persists at all seasons and warms the high-heat-capacity atmosphere which increases the thermal inertia of the system.  In this ``moist'' case, the convective updraft of the Hadley circulation is in quadrature with solar forcing, which translates to a $\sim$ 7 year time lag.  If instead surface evaporation is uniformly restricted, the numerical model has the appropriate amount of time lag, matching the observed lag in the spin rate drift (see Figure \ref{fig2}).  Our model predicts the spin rate drift has changed directions; current Cassini RADAR observations of Titan may show this trend.

In conclusion, the observed time lag of Titan's drift in rotation rate is consistent with the lag introduced by thermodynamic-dynamic coupling resulting from the methane cycle in the atmosphere of Titan.  We have shown the change in total atmospheric angular momentum is primarily controlled by the latitude of the Hadley cell updraft, and this updraft marks the location of the most persistent clouds features.  The recent disappearance of persistent south polar clouds \citep{Schaller.et.al.06b} suggests the Hadley cell updraft is moving equatorward in response to the changing season.  The resulting transfer of angular momentum to the atmosphere from the surface should currently be decelerating the spin rate, as shown in Figure \ref{fig2}.  Other sources of torque might measurably contribute to the observed spin rate drift; however better knowledge of Titan's gravitational moments and interior structure is required to estimate their effects.

\begin{acknowledgements}
I wish to thank Peter Goldreich for helpful discussions.  I gratefully acknowledge support from the Institute for Advanced Study and the W. M. Keck Foundation Fund.  
\end{acknowledgements}

\end{document}